\documentclass[prd,onecolumn,nofootinbib]{revtex4}%
\usepackage{graphicx}
\usepackage{amsmath}
\usepackage{amsfonts}
\usepackage{amssymb}
\usepackage{color}%
\usepackage{dcolumn}
\providecommand{\U}[1]{\protect\rule{.1in}{.1in}}

\newcommand{\f}{\begin{equation}}
\newcommand{\ff}{\end{equation}}
\newcommand{\fa}{\begin{eqnarray}}
\newcommand{\ffa}{\end{eqnarray}}

\begin{document}
\title{A note on superluminal neutrinos and deformed special relativity }

\author{Yi Ling $^{1,2}$}
\email{yling@ncu.edu.cn}

\affiliation{$^1$ Institute of High Energy Physics, Chinese
Academy of Sciences, Beijing 100049, China\\ $^2$ Center for
Relativistic Astrophysics and High Energy Physics, Department of
Physics, Nanchang University, 330031, China}

\begin{abstract}
In this short note we remark that the OPERA measurement could be
reconciled with information from SN1987a in the context of
deformed special relativity without the loss of energy through
Cherenkov-like process reported by Cohen and Glashow.
\end{abstract} \maketitle

Soon after the OPERA collaboration reported the evidence of the
superluminal behavior for neutrinos with energies at the GeV
level\cite{2011zb}, Cohen and Glashow argued that this observation
conflicts with known physics \cite{Cohen2011hx}( some other
relevant work can also be found in
\cite{Gonzalez11jc,Bi11nd,Li11ad}). The key point in their
argument is that these neutrinos would loose much of their energy
via Cherenkov-like processes since the threshold energy of the
decaying interaction would decease down to about 140 MeV due to
the superluminality of neutrinos. In their framework the Lorentz
symmetry is manifestly broken and the presence of a preferred
frame is implied. On the contrary, vert recently Amelino-Camelia
$et. al.$ show in \cite{Amelino2011bz}that this confliction can be
overcome in the framework of deformed special relativity, in which
the Lorentz symmetry is not completely broken but deformed such
that the modified form of the dispersion relation is still
invariant under the Lorentz boost. They presented a specific
example to demonstrate that the anomalous process is forbidden in
the deformed-Lorentz-symmetry case. Therefore, given that the
principle of relativity of inertial frames is preserved, in
principle the superluminality of neutrinos detected by the OPERA
need not contradict known theory of physics. Unfortunately, as the
authors themselves claimed in \cite{Amelino2011bz} that this
example with a specific modified dispersion relation can not
easily fit the data on both superluminal neutrinos and SN1987a. In
this short note we intend to remark that this framework can
actually be extended to have a general form of dispersion relation
such that in principle the OPERA measurement could be reconciled
with information from SN1987a. Without loss of generality, we
assume that the deformed dispersion relation has the following
form
 \f f^2(p)E^2-p^2=m^2,
\ff where $f(p)$ is a function of the momentum constrained by the
condition that $f(p)\rightarrow 1$ as $p\rightarrow 0$. The
invariance of this deformed dispersion relation under the Lorentz
boost requires that  \f [N_i, f^2(p)E^2-p^2]=0, \ff where $N_i$ is
the generator of the boost and in general may be deformed with a
dependence on the mass of particle. With a simple calculation, one
can show that the invariance can be achieved by the following
deformed algebra
\fa [N_i, p_j] &=& f(p)E\delta_{ij},\nonumber\\
\left[ N_i, E\right]  &=& p_i\left( \frac{1}{
f(p)}-2E^2\frac{\partial f(p)}{\partial p^2}\right). \ffa Now let
us consider the same process as presented in \cite{Amelino2011bz},
namely a single incoming particle with light mass $m$ and
energy-momentum $(E,p)$ decaying into two outgoing particles with
the same mass $M$, but energy-momentum $(E_1,p_1)$ and $(E_2,p_2)$
respectively. Then the conservation laws for momentum and energy
in the context of deformed special relativity are read as
 \fa \vec{p}&=&\vec{p}_1+\vec{p}_2\nonumber\\
f(p)E &=& f(p_1)E_1+ f(p_2)E_2.\ffa Similarly, suppose the opening
angle of the directions of the outgoing particles is $\theta$,
then one can obtain the following result under the
ultra-relativistic limit with $E\gg m, E_1\gg M, E_2\gg M$ \f
\cos\theta=\frac{2f(p_1)f(p_2)E_1E_2-m^2+2M^2}{2f(p_1)f(p_2)E_1E_2-
M^2\left(\frac{f(p_1)E_1}{f(p_2)E_2}+\frac{f(p_2)E_2}{f(p_1)E_1}\right)}.
\ff Since $m<M$, it is obvious that $cos\theta$ is always greater
than one whatever the specific form of $f(p)$ is, implying that
the anomalous process is always forbidden. It means that in the
context of deformed special relativity one is free to choose the
function $f(p)$ to fit the data, needless to worry about the
adjustment of $cos\theta$ due to the change of the dispersion
relation. For instance, if we show our respect to the observed
data in both OPERA and FERMILAB1979, then the superluminality of
neutrinos is expected to be of the same order with $\delta c\sim
10^{-5}$ in the energy interval from $3 GeV$ to $200
GeV$\cite{Amelino2011dx,Giudice2011mm}. One may propose a deformed
dispersion relation with $f^2(p)$ as  \f
f^2(p)=1-\frac{2p^2}{mM_p+\beta p^2},\ff where $m$ is the mass of
neutrinos with a magnitude of $10^{-3}ev$, and $\beta$ is a new
parameter introduced with an order as $\beta\sim 10^5$, which
might arise as the ratio of the fundamental(Planck) scale $M_p$ to
some characteristic scale $\bar{M}$ which is a little below the
GUT scale. For this deformed dispersion relation , it is easy to
see that as $mM_p\gg\beta p^2$ \f f^2(p)\simeq
1-\frac{2p^2}{mM_p},\label{dsa}\ff such that $\delta c$ can climb
up quickly from a tiny value in the energy region of SN1987a to a
relative large number when reaching the energy region detected by
OPERA. Specifically, when the energy of neutrinos is up to
$10GeV$, the second term has an order of $10^{-5}$ which gives
rise to a superluminality consistent with the OPERA data. The form
in Eq.(\ref{dsa}) was originally proposed in \cite{Magueijo11xy}
as a fit to the OPERA data. Unfortunately, when the energy
increases further, say to $100GeV$, the superluminality based on
this dispersion relation is obviously much larger and incompatible
with the observed data in FERMILAB1979. However, for our proposed
dispersion relation, when $mM_p\ll\beta p^2$ \f f^2(p)\simeq
(1-\frac{2}{\beta})+\frac{2mM_p}{\beta^2p^2}, \ff such that the
superluminality is of the same order as $1/\beta$ and the higher
order corrections are greatly suppressed by the energy of
particles.

Inspired by the analysis in \cite{Giudice2011mm}, the best fitting
with both OPERA data and SN1987a data  may be achieved by a more
general dispersion relation as \f
f^2(p)=1-\frac{2p^n}{(mM_p)^{n/2}+\beta p^n},\ff with $n>2$.

\begin{acknowledgments}
I would like to thank Wei-Jia Li and Jian-Pin Wu for very helpful
discussion during the early stage of this work. I am also grateful
to G.~Amelino-Camelia, L.~Freidel, J.~Kowalski-Glikman and
L.~Smolin for correspondence. This work is partly supported by
NSFC(10875057), Fok Ying Tung Education Foundation(No. 111008),
the key project of Chinese Ministry of Education(No.208072),
Jiangxi young scientists(JingGang Star) program and 555 talent
project of Jiangxi Province. We also acknowledge the support by
the Program for Innovative Research Team of Nanchang University.
\end{acknowledgments}

\end{document}